\documentclass[conference]{IEEEtran}
\IEEEoverridecommandlockouts

\usepackage[T1]{fontenc}
\usepackage{cite}
\usepackage{amsmath,amssymb,amsfonts}
\usepackage{algorithmic}
\usepackage{graphicx}
\usepackage{textcomp}
\usepackage[table]{xcolor}
\usepackage{array}
\usepackage{tabularx}
\usepackage{booktabs}
\usepackage{multirow}
\usepackage{url}
\usepackage{tikz}
\usetikzlibrary{arrows.meta}
\usepackage[printonlyused,nolist]{acronym}

\graphicspath{{figures/}}
\newcommand\copyrighttext{%
  \footnotesize \textcopyright \the\year{} This work has been submitted to the IEEE for possible publication. Copyright may be transferred without notice, after which this version may no longer be accessible}

\newcommand\copyrightnotice{%
\begin{tikzpicture}[remember picture,overlay]
\node[anchor=north,yshift=0pt] at (current page.north) {\fbox{\parbox{\dimexpr0.9\textwidth-\fboxsep-\fboxrule\relax}{\copyrighttext}}};
\end{tikzpicture}%
}

\begin{document}
\bstctlcite{IEEEexample:BSTcontrol}
\begin{acronym}
    \acro{AI}{Artificial Intelligence}
    \acro{ML}{Machine Learning}
    \acro{CNN}{Convolutional Neural Network}
    \acro{FC}{Fully Connected Layers}
    \acro{5G-NR}[5G NR]{fifth-generation New Radio}
    \acro{5G}{fifth-generation}
    \acro{DL}{deep learning}
    \acro{ReLU}{Rectified Linear Unit}
    \acro{DNN}{Deep Neural Network}
    \acro{ONNX}{Open Neural Network Exchange}
    \acro{GPU}{Graphics Processing Unit}
    \acro{CPU}{Central Processing Unit}
    \acro{SRCNN}{Super Resolution Convolutional Neural Network}
    \acro{EDSR}{Enhanced Deep Super-Resolution}
    \acro{ViT}{Vision Transformer}
    \acro{SR}{super-resolution}
    \acro{IR}{Image Restoration}
    \acro{LR}{Low-Resolution}
    \acro{LS}{least squares}
    \acro{LI}{linear interpolation}
    \acro{HR}{High-Resolution}
    \acro{MDSR}{Multi-Scale Approach}
    \acro{MAE}{Mean Absolute Error}
    \acro{CE}{channel estimation}
    \acro{NMSE}{normalized mean-squared error}
    \acro{MSE}{mean-squared error}
    \acro{MHSA}{multi-head self-attention}
    \acro{BER}{Bit Error Rate}
    \acro{CSI}{Channel State Information}
    \acro{3GPP}{3rd Generation Partnership Project}
    \acro{OFDM}{orthogonal frequency-division multiplexing}
    \acro{FFT}{fast Fourier transform}
    \acro{SNR}{signal-to-noise ratio}
    \acro{AWGN}{additive white Gaussian noise}
    \acro{SISO}{single-input single-output}
    \acro{I/Q}{In-Phase and Quadrature Components}
    \acro{1D}{One-Dimensional}
    \acro{2D}{two-dimensional}
    \acro{AR}{Augmented Reality}
    \acro{VR}{Virtual Reality}
    \acro{UHD}{Ultra-High-Definition}
    \acro{TDL}{tapped delay line}
    \acro{TTI}{transmission time interval}
    \acro{DnCNN}{Denoising Convolutional Neural Network}
    \acro{Conv}{Convolutional}
    \acro{NR}{New Radio}
    \acro{MMSE}{minimum mean-squared error}
    \acro{LMMSE}{linear minimum mean-squared error}
    \acro{DRN}{Deep Residual Networks}
    \acro{ResNet}{Residual Network}
    \acro{RB}{resource block}
    \acro{SCS}{subcarrier spacing}
    \acro{CP}{cyclic prefix}
    \acro{PDSCH}{Physical Downlink Shared Channel}
    \acro{SR-Net}{Super-Resolution Network}
    \acro{CIR}{Channel Impulse Response} 
    \acro{SE}{squeeze-and-excitation}
    \acro{ECA}{Efficient Channel Attention}
    \acro{GAP}{Global Average Pooling}
    \acro{SRDnNet}{Super Resolution De-noising Convolutional Neural Network}
    \acro{ReEsNet}{Residual channel Estimation Network}
    \acro{UE}{user equipment}
    \acroplural{UE}[UEs]{user equipment devices}
    \acro{IQ}{In-phase an Quadrature}
    \acro{HELENA}{High-Efficiency Learning-based channel Estimation using dual Neural Attention}
    \acro{FLOPS}{Floating-point operations per second}
    \acro{Conv2D}{2D Convolutional}
    \acro{MBConv}{Mobile Inverted Bottleneck Convolution}
    \acro{HARQ}{Hybrid Automatic Repeat Request}
    \acro{FPGA}{Field-Programmable Gate Array}
    \acro{LSiDNN}{LS-augmented interpolated Deep Neural
Network}
    \acro{AttRNet}{Attention mechanism and Residual
Network}
    \acro{CEViT}[CE-ViT]{Channel Estimator Vision Transformer}
    \acro{PD-CEViT}{pilot-design channel estimator vision transformer}
    \acro{OTFS}{Orthogonal Time Frequency Space}
    \acro{UAV}{Unmanned Aerial Vehicle}
    \acro{LoS}{Line-of-Sight}
    \acro{NLoS}{Non-Line-of-Sight}
    \acro{NDT}{Network Digital Twin}
    \acro{MIMO}{Multiple-Input Multiple-Output}
    \acro{dB}{Decibels}
    \acro{ProEsNet}{Progressive Estimation Network}
    \acro{EPformer}{Efficient Parallel Transformer}
    \acro{NTN}{non-terrestrial network}
    \acro{TN}{Terrestrial Network}
    \acro{LEO}{low Earth orbit}
    \acro{NGSO}{Non-Geostationary Satellite Orbit}
    \acro{CFO}{carrier-frequency offset}
    \acro{DM-RS}{Demodulation Reference Signal}
    \acro{PA-LS}{Pilot-Aided Least Squares}
    \acro{DA-LS}{Data-Aided Least Squares}
    \acro{MDELAN}{Multi-Dilated Efficient Layer Aggregation Network}
    \acro{TPE}{Tree-structured Parzen Estimator}
    \acro{P99}{99th-percentile}
    \acro{QC}{quality control}
    \acro{FP16}{16-bit floating point}
    \acro{FP32}{32-bit floating point}
    \acro{INT8}{8-bit integer}
    \acro{GNSS}{global navigation satellite system}
    \acro{SIB19}{System Information Block 19}
    \acro{ICI}{inter-carrier interference}
    \acro{RE}{resource element}
    \acro{CDM}{code-division multiplexing}
    \acro{QPSK}{quadrature phase-shift keying}
    \acro{SSB}{synchronization signal block}
    \acro{PBCH}{physical broadcast channel}
    \acro{BLER}{block error rate}
    \acro{COTS}{commercial off-the-shelf}
    \acro{FLOPs}{floating-point operations}
    \acro{ANNA}{Attention-free Neural Neighborhood Aggregation}
    \acro{NRC}{No Residual Connection}
    \acro{CHEA}{Channel Estimation Attention}
    \acro{MAC}{multiply--accumulate operation}
    \acro{UL}{uplink}
    \acro{PUSCH}{Physical Uplink Shared Channel}
\end{acronym}

\title{ANNA: Attention-Free Neural Neighborhood Aggregation for Channel Estimation on Regenerative NTN Payloads
\thanks{This work was supported by the imec.icon RAPIDNESS project,
co-financed by imec and Flanders Innovation \& Entrepreneurship
(HBC.2024.0772).}
}

\author{
\IEEEauthorblockN{Miguel Camelo Botero, Nina Slamnik-Krije\v{s}torac,
and Johann Marquez-Barja}
\IEEEauthorblockA{\textit{IDLab, University of Antwerp -- imec}}
}


\maketitle

\copyrightnotice
\begin{abstract}
Accurate \ac{DL}-based \ac{CE} for \ac{NTN} uplinks must meet the processing, power, and latency constraints of regenerative payloads. We propose \ac{ANNA}, a lightweight architecture combining per-\ac{RB} pilot projection, convolutional neighborhood aggregation, and shared reconstruction, with dense-grid expansion only at the output. Evaluation against learned and classical estimators uses paired waveform-generated datasets with receiver frequency correction enabled and disabled. With correction enabled, \ac{FP16} deployment on a 10~W Jetson Orin NX achieves 0.372~ms \ac{P99} latency and 2.15~\acs{dB} lower mean \ac{NMSE} than the fastest learned baseline at 29.6\% higher latency. Relative to the most accurate learned baseline, \acs{ANNA} is 4.08$\times$ faster and consumes 80.5\% less module energy per sample, with a 1.14~\acs{dB} accuracy penalty. It meets the adopted 0.5~ms \acs{P99} target in all 100 benchmark runs, demonstrating a favorable accuracy--latency--energy trade-off for embedded uplink \ac{CE}.
\end{abstract}


\acresetall
\IEEEPARstart{I}{n} \ac{LEO} \acp{NTN}, regenerative payloads can host some or all base-station processing~\cite{TR38821,Petry2024Regenerative}. Their onboard \ac{UL} receivers must estimate the channel under residual Doppler and synchronization impairments within processing and power constraints~\cite{TR38811}. Conventional pilot-based \ac{LS} estimation is inexpensive but noise-sensitive, whereas \ac{LMMSE} requires channel statistics and matrix processing~\cite{Edfors1998}. These limitations motivate \ac{DL}-based estimators that learn channel reconstruction from representative data \cite{soltani2019deep, HELENA2025, CHEA2026}.

Terrestrial \ac{DL}-based \ac{CE} includes convolutional and residual networks~\cite{soltani2019deep,maruyama2021,Gao2025,Zhang2024}, fully connected models~\cite{Sharma2024}, and attention-based architectures such as \ac{CEViT}, \ac{HELENA}, and \ac{CHEA}~\cite{CEViT2023,HELENA2025,CHEA2026}. For \ac{NTN} uplinks, \ac{MDELAN} instead builds on the \ac{CE} stage of the model-driven MDX receiver~\cite{MDX2025,MDELAN2026}. These approaches offer different accuracy--cost trade-offs.

Our recent downlink study showed that terrestrial estimators remain effective after \ac{NTN} retraining without architectural changes \cite{HELENANTN2027}. \ac{HELENA} provided a favorable accuracy--latency--energy trade-off and met the adopted 0.5~ms \ac{P99} budget on a high-end \ac{GPU}, but exceeded it on a 10~W Jetson Orin NX~\cite{HELENANTN2027}. This embedded limitation is relevant to regenerative payloads, whose proposed designs include Orin for onboard \ac{AI} workloads~\cite{Benoot2025VersalOrin}.

Motivated by these limitations, we first investigate whether simplifying \ac{HELENA} can retain its accuracy while meeting the latency target. We consider reducing attention capacity and replacing full-grid features with compact pilot inputs, motivated by designs such as \ac{LSiDNN} and \ac{CHEA}~\cite{Sharma2024,CHEA2026}. As the results will show, attention reduction alone leaves latency above target, while compact-input variants incur appreciable accuracy loss. These findings motivate \ac{ANNA}, a novel lightweight architecture that jointly designs pilot representation and reconstruction. It projects pilots within each \ac{RB}, aggregates neighboring features through convolutions, and reconstructs channel patches through a shared linear dictionary. This attention-free design exploits local frequency structure while deferring dense-grid expansion until the output.

Building on this design, this paper makes three main contributions: a) \acs{ANNA}, a lightweight architecture for low-latency uplink \ac{CE}; b) paired waveform-generated \ac{NTN} \ac{UL} datasets with receiver frequency correction enabled (UL-\acs{NTN}-1) and disabled (UL-\acs{NTN}-2), complementing the downlink datasets in~\cite{HELENANTN2027}; and c) a unified benchmark of 28 \ac{CE} configurations, comprising 6 \acs{ANNA} and 9 \acs{HELENA} variants, 9 other \ac{DL}-based estimators, and 4 classical estimators, evaluated under both receiver conditions, with learned-model latency and energy measured on a 10~W Jetson Orin NX using UL-\acs{NTN}-1.


\section{System Model and Problem Statement}\label{sec:system}
\begin{figure}[t]
    \centering
    \includegraphics[width=\columnwidth]{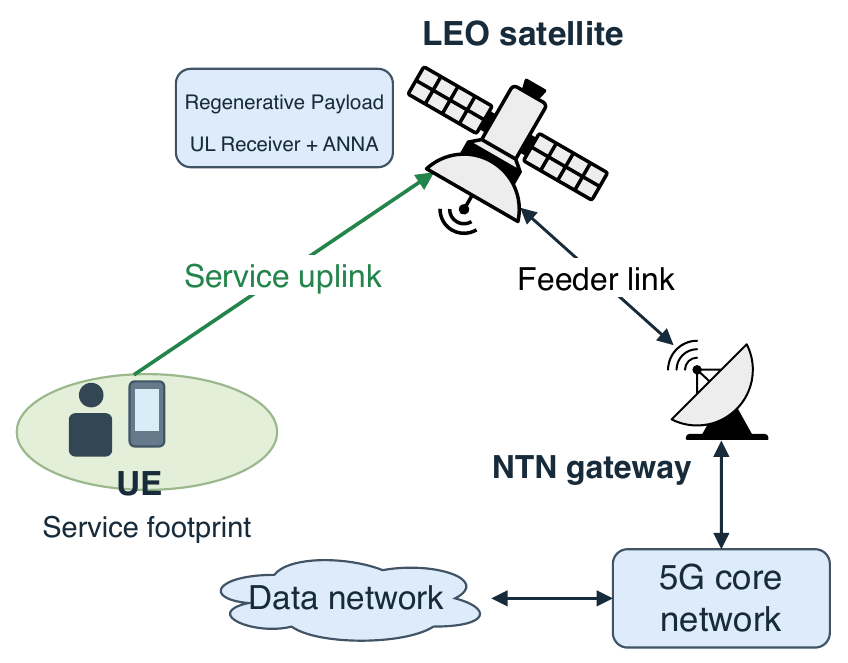}
    \caption{Regenerative \acs{NTN} uplink with onboard \acs{CE}.}
    \label{fig:uplink_system}
\end{figure}

We consider a \ac{SISO} \ac{5G-NR} service uplink from a terrestrial \ac{UE} to a \ac{LEO} satellite carrying a regenerative payload, with onboard reception and \ac{CE}. The \ac{PUSCH} uses \ac{OFDM} with a \ac{CP} and no transform precoding. Figure~\ref{fig:uplink_system} shows the service uplink and connection to the terrestrial core network. Bold uppercase and lowercase symbols denote matrices and vectors, respectively, and $j=\sqrt{-1}$.

Let $\mathbf X\in\mathbb C^{N_f\times N_t}$ be the transmitted grid, with $N_f=12B$ subcarriers across $B$ allocated \acp{RB} and $N_t$ \ac{OFDM} symbols. Indices $i\in\{1,\ldots,N_f\}$ and $k\in\{1,\ldots,N_t\}$ identify subcarriers and symbols, respectively. Each allocated position contains nonzero data or \ac{DM-RS} symbols; only the latter are known to the pilot-based estimator.

Let $f_{\mathrm{sat}}$ denote the signed satellite-induced Doppler shift, assumed constant over one transmitted slot, and let $\hat f_{\mathrm{UE}}$ be the \ac{UE}'s estimate of this shift for transmitter pre-compensation, both in hertz. The estimate uses satellite ephemeris and the \ac{UE} position~\cite{TR38821}. Unlike the downlink beam-reference correction in~\cite{HELENANTN2027}, which removes beam-common Doppler, the uplink correction targets the Doppler of the individual \ac{UE}--satellite link. With sampling interval $T_s$ in seconds and sample index $n$, pre-compensation maps the \ac{OFDM} waveform $x[n]$ into
\begin{equation}
x_{\mathrm{pc}}[n]=x[n]e^{-j2\pi\hat f_{\mathrm{UE}}nT_s},
\label{eq:ul_precompensation}
\end{equation}
where the subscript $\mathrm{pc}$ identifies transmitter pre-compensation. The residual \ac{CFO} at the receiver input is
\begin{equation}
f_{\mathrm{res}}=f_{\mathrm{sat}}-\hat f_{\mathrm{UE}}+f_{\mathrm{osc}}
=e_{\mathrm{pc}}+f_{\mathrm{osc}},
\label{eq:ul_residual_frequency}
\end{equation}
where $e_{\mathrm{pc}}=f_{\mathrm{sat}}-\hat f_{\mathrm{UE}}$ is the Doppler pre-compensation error, and $f_{\mathrm{osc}}$ is the net transmitter--receiver oscillator offset. \ac{UE} mobility is represented separately by the Doppler spread of the time-varying multipath channel.

Let $\mathcal C_{f_{\mathrm{sat}}}\{\cdot\}$ denote the channel operator, including multipath propagation, mobility-induced time variation, and satellite Doppler. The received waveform is
\begin{equation}
y[n]=e^{j2\pi f_{\mathrm{osc}}nT_s}
\mathcal C_{f_{\mathrm{sat}}}\{x_{\mathrm{pc}}\}[n]+w[n],
\label{eq:ul_received_waveform}
\end{equation}
where $w[n]\sim\mathcal{CN}(0,\sigma_w^2)$ is independent, zero-mean complex Gaussian noise with $\mathbb E[|w[n]|^2]=\sigma_w^2$. The noise-free received waveform is $y^{\mathrm{clean}}[n]=y[n]-w[n]$. Frequency offsets and channel time variation act before \ac{OFDM} demodulation and can therefore produce \ac{ICI}.

The onboard receiver estimates timing from the received \ac{DM-RS} and may compensate the residual frequency offset using \ac{CP} redundancy and \ac{DM-RS}\footnote{\label{fn:matlab_pusch}\url{https://www.mathworks.com/help/5g/ug/nr-ntn-pusch-throughput.html}}. Let $s$ index the receiver-processing configuration and $\hat f_{\mathrm{RX},s}$ denote its applied frequency correction. The remaining offset is
\begin{equation}
f_{\mathrm{post},s}=f_{\mathrm{res}}-\hat f_{\mathrm{RX},s},
\label{eq:ul_post_frequency}
\end{equation}
where $\mathrm{RX}$ denotes receiver-side correction and $\mathrm{post}$ the offset remaining after correction.

Let $\mathcal R_s(\cdot;\hat{\boldsymbol\eta}_s)$ apply timing alignment, the selected frequency correction, and \ac{OFDM} demodulation, where $\hat{\boldsymbol\eta}_s$ contains the timing estimates and applied frequency correction obtained from $y[n]$. The received grid is $\mathbf Y^{(s)}=\mathcal R_s(y;\hat{\boldsymbol\eta}_s)\in\mathbb C^{N_f\times N_t}$, with entries $Y_{i,k}^{(s)}$. As in~\cite{HELENANTN2027}, applying the same synchronization estimates to the noise-free received waveform defines the effective channel target
\begin{equation}
\mathbf H_s^{\mathrm{eff}}=
\mathcal R_s(y^{\mathrm{clean}};\hat{\boldsymbol\eta}_s)
\oslash\mathbf X,
\label{eq:effective_label}
\end{equation}
where $\oslash$ denotes element-wise division and $\mathrm{eff}$ identifies the effective channel in the receiver reference frame. Under residual \ac{ICI}, the target depends on the transmitted grid rather than solely on the propagation channel.

We consider two receiver variations: UL-\acs{NTN}-1 ($s=1$) applies the estimated residual-frequency correction, whereas UL-\acs{NTN}-2 ($s=2$) disables it by setting $\hat f_{\mathrm{RX},2}=0$. Both retain transmitter pre-compensation and receiver timing alignment. Matching transmitted waveforms, channels, and noise realizations isolates the effect of receiver frequency correction.

For either configuration, let $\mathcal P$ denote the set of \ac{DM-RS} positions. The pilot \ac{LS} estimate is
\begin{equation}
\hat H_{i,k}^{\mathrm{LS},(s)}=
Y_{i,k}^{(s)}/X_{i,k},\qquad (i,k)\in\mathcal P.
\label{eq:ul_pilot_ls}
\end{equation}
Let $\mathbf H_{\mathrm{in},s}$ denote a model-specific representation of these observations: a zero-filled pilot \ac{LS} grid, an interpolated grid, or a compact arrangement of pilot coefficients. With optional auxiliary features $\mathbf u_s$, such as pilot masks, side information, or data-aided estimates, a learned estimator implements
\begin{equation}
\hat{\mathbf H}_s^{\mathrm{eff}}=f_{\Theta_s}(\mathbf H_{\mathrm{in},s},\mathbf u_s),
\label{eq:ul_estimator}
\end{equation}
where $\Theta_s$ contains the parameters trained for receiver condition $s$; $\mathbf u_s$ is omitted when not required. \emph{The \ac{CE} problem is to reconstruct $\mathbf H_s^{\mathrm{eff}}$ over the full allocated grid from pilot-based observations, minimizing estimation error while meeting the receiver's inference-latency and computational-resource constraints.}

\section{ANNA Architecture}\label{sec:architecture}
\label{sec:anna_rationale}
Reducing the inference cost of neural \ac{CE} requires jointly considering input representation and reconstruction. As shown in Section~\ref{sec:results}, reducing \acs{HELENA}'s attention capacity alone does not meet the adopted latency target, while replacing its full-grid input with packed pilots incurs an accuracy penalty. \ac{ANNA} addresses these limitations jointly: it retains compact pilot observations and uses their local frequency structure to support reconstruction with a lightweight network.

\begin{figure}[tb]
    \centering
    \includegraphics[width=\columnwidth]{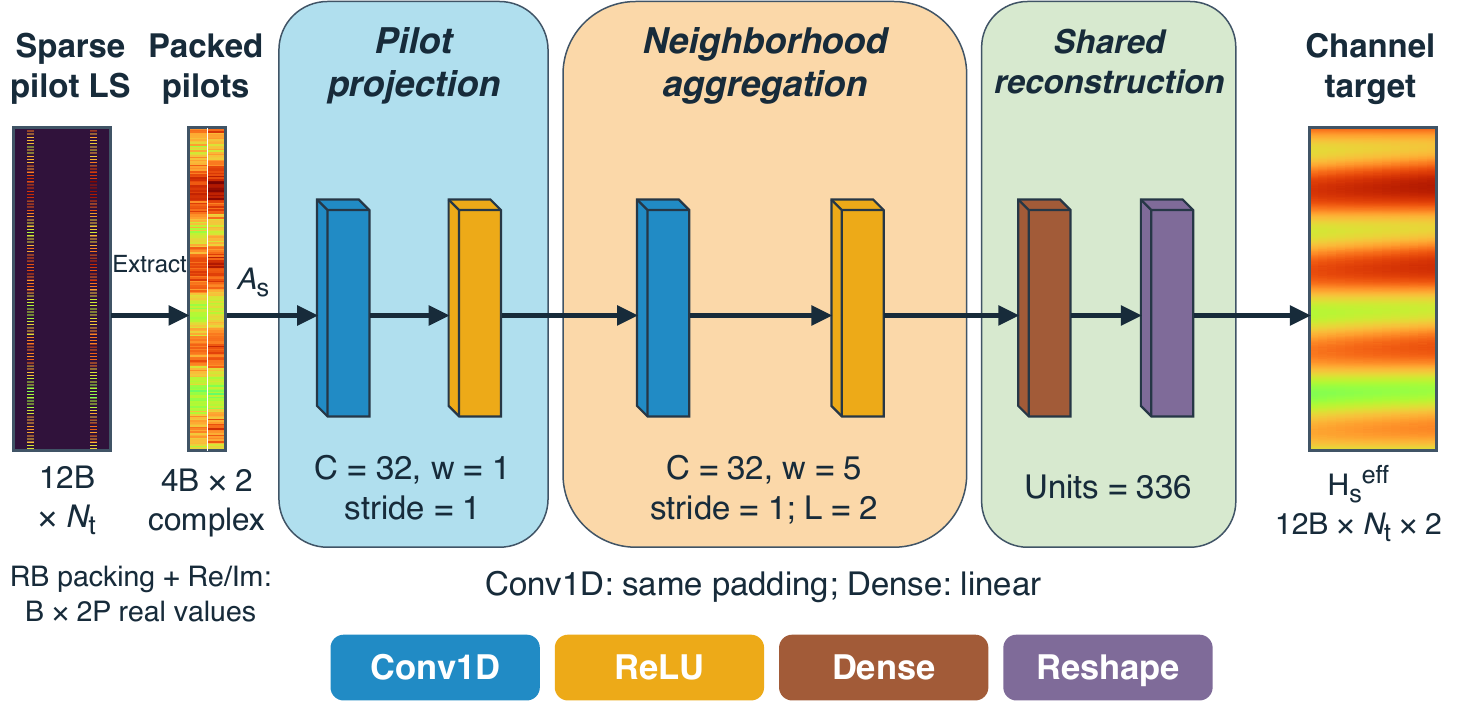}
    \caption{\acs{ANNA}: pilot projection, neighborhood aggregation, and shared reconstruction.}
    \label{fig:anna_architecture}
\end{figure}

As shown in Fig.~\ref{fig:anna_architecture}, \acs{ANNA} first encodes the pilots within each \ac{RB}, then combines features from neighboring \acp{RB}, and finally reconstructs the channel through a shared linear map. Nonlinear processing remains in the compact pilot domain until output expansion.

\subsection{Pilot Representation and Feature Projection}
Unlike the full-grid inputs of \acs{HELENA} and \acs{CEViT}, \acs{ANNA} packs pilot \ac{LS} coefficients into $\mathbf A_s\in\mathbb R^{B\times 2P}$, where $B$ is the number of allocated \acp{RB}, $P$ the number of complex pilot coefficients per \ac{RB}, and $N_t$ the number of \ac{OFDM} symbols in the grid.

Each row stores one \ac{RB}'s real and imaginary pilot components in a fixed order. The full grid stores $12B\times N_t\times2=24BN_t$ real values, since each \ac{NR} \ac{RB} contains 12 subcarriers with complex coefficients. Packing preserves the pilot \ac{LS} coefficients in $2BP$ real values. For $B=51$, $P=8$, and $N_t=14$, this is a 21-fold reduction. Thus, in~\eqref{eq:ul_estimator}, $\mathbf H_{\mathrm{in},s}=\mathbf A_s$, without auxiliary features $\mathbf u_s$, such as the nominal \ac{SNR}, mobility Doppler, and delay spread used by \acs{CEViT}.

The first stage transforms each \ac{RB}'s $2P$ real pilot values into $C$ learned features. Because every row of $\mathbf A_s$ represents the same within-\ac{RB} pilot positions, this projection uses the same weights across \acp{RB}. We omit the receiver-configuration index $s$ from intermediate features and weights, retaining separately trained parameters $\Theta_s$ for each condition. A width-one convolution followed by \ac{ReLU} implements the projection without mixing neighboring \acp{RB}:
\begin{equation}
\mathbf Z^{(0)}=\rho\!\left(\mathbf A_s\mathbf W_{\mathrm{in}}
+\mathbf 1_B\mathbf b_{\mathrm{in}}^{\mathsf T}\right),
\label{eq:anna_projection}
\end{equation}
where $\mathbf W_{\mathrm{in}}\in\mathbb R^{2P\times C}$, $\mathbf b_{\mathrm{in}}\in\mathbb R^C$, $\mathbf 1_B\in\mathbb R^{B\times1}$ contains ones, and $\rho(u)=\max(0,u)$ acts element-wise. The output $\mathbf Z^{(0)}\in\mathbb R^{B\times C}$ retains one feature vector per \ac{RB}. Temporal pilot positions occupy distinct feature coordinates, allowing the projection to combine them without constructing a full-grid feature map.

\subsection{Neighborhood Aggregation}
Frequency correlation in multipath channels makes neighboring pilot observations useful for denoising and reconstruction~\cite{Edfors1998}. \acs{ANNA} exploits this structure through one-dimensional convolutions along the \ac{RB} axis. Shared kernels combine features from neighboring \acp{RB} without computing allocation-wide attention. Each aggregation layer applies a convolution followed by \ac{ReLU}, without a residual addition.

Let $\mathbf z_b^{(\ell-1)}\in\mathbb R^{1\times C}$ denote row $b\in\{1,\ldots,B\}$ entering aggregation layer $\ell\in\{1,\ldots,L\}$. For odd kernel width $w$ and radius $r=(w-1)/2$,
\begin{equation}
\mathbf u_b^{(\ell)}=
\sum_{q=-r}^{r}\mathbf z_{b+q}^{(\ell-1)}\mathbf W_q^{(\ell)}
+(\mathbf b_{\mathrm{sp}}^{(\ell)})^{\mathsf T},
\label{eq:anna_spatial}
\end{equation}
where $\mathbf W_q^{(\ell)}\in\mathbb R^{C\times C}$ and $\mathbf b_{\mathrm{sp}}^{(\ell)}\in\mathbb R^C$ are learned weights and biases. Out-of-allocation features are zero-padded. Stacking the rows gives $\mathbf U^{(\ell)}\in\mathbb R^{B\times C}$, and activation yields
\begin{equation}
\mathbf Z^{(\ell)}=\rho\!\left(\mathbf U^{(\ell)}\right).
\label{eq:anna_single_layer_block}
\end{equation}
Each layer therefore combines features from neighboring \acp{RB} while preserving the compact $B\times C$ layout. With unit stride and no dilation, $L$ layers have a receptive field of $1+L(w-1)$ \acp{RB}, truncated at the allocation boundaries. Kernel width controls frequency context; the feature channels encode pilot observations from both pilot-bearing symbols.

\subsection{Shared Reconstruction Dictionary}\label{sec:anna_dictionary}
The final stage reconstructs the effective channel in~\eqref{eq:effective_label}. Each \ac{RB} requires the same output coordinates: 12 subcarriers over $N_t$ symbols, with real and imaginary components. We therefore use a shared linear map to reconstruct each channel patch from its neighborhood features. With $D=24N_t$ real values per patch, reconstruction follows
\begin{equation}
\mathbf Q_s=\mathbf Z^{(L)}\mathbf W_{\mathrm{out}}
+\mathbf 1_B\mathbf b_{\mathrm{out}}^{\mathsf T},
\label{eq:anna_reconstruction}
\end{equation}
where $\mathbf W_{\mathrm{out}}\in\mathbb R^{C\times D}$ and $\mathbf b_{\mathrm{out}}\in\mathbb R^D$. The rows of $\mathbf W_{\mathrm{out}}$ form a learned dictionary of patch patterns, combined using the features inferred for each \ac{RB}. Thus, the reconstruction weights are shared across the allocation, while the estimated patches vary with their local pilot observations. Reshaping $\mathbf Q_s\in\mathbb R^{B\times D}$ into $12B\times N_t\times2$ and combining its real and imaginary components produces $\hat{\mathbf H}_s^{\mathrm{eff}}$.

\section{Datasets and Experimental Setup}\label{sec:dataset}
\subsection{Uplink Datasets}
UL-\acs{NTN}-1 and UL-\acs{NTN}-2 extend the paired waveform-generation approach of our downlink study~\cite{HELENANTN2027} to \ac{PUSCH} reception at a regenerative payload. Generation uses MATLAB 5G Toolbox and Satellite Communications Toolbox, adapting the regenerative \ac{NTN} uplink example\textsuperscript{\ref{fn:matlab_pusch}}. The configuration used for the datasets is listed in Table~\ref{tab:ul_dataset}.

Unlike the downlink setup, the uplink transmitter applies \ac{UE}-specific Doppler pre-compensation. Independent pre-compensation error and oscillator offset leave a receiver-input residual in $[-400,400]$~Hz. The waveform contains no phase-tracking reference signals; its \ac{DM-RS} pilots occupy symbols 2 and 11 and within-\ac{RB} subcarriers $\{0,1,6,7\}$, using zero-based indexing. These positions provide eight complex pilot coefficients per \ac{RB}, from which the estimators reconstruct the effective-channel target in~\eqref{eq:effective_label}.

Each channel profile contains 256 scenarios, partitioned into 179 training, 38 validation, and 39 test scenarios before generating the 11 \ac{SNR} variants. Across the four NTN channel profiles, this yields 7,876 training, 1,672 validation, and 1,716 test samples. Realizations are matched across receiver conditions, with independent noise across \acp{SNR}.

\begin{table}[tb]
\centering
\caption{Paired regenerative-uplink dataset configuration.}
\label{tab:ul_dataset}
\setlength{\tabcolsep}{3pt}
\renewcommand{\arraystretch}{1.05}
\begin{tabular}{lp{5.1cm}}
\toprule
Parameter & Common configuration\\
\midrule
Carrier/grid & 2 GHz; 51 \acsp{RB}; 30 kHz \acs{SCS}; normal \acs{CP}\\
Grid/link & $612\times14$ complex coefficients; \acs{SISO} \acs{PUSCH}\\
Modulation & Uncoded \acs{QPSK}; no transform precoding\\
\acs{DM-RS} & Mapping A; position 2; additional position 1; configuration type 2; one CDM group\\
Pilot count & 408 complex coefficients; eight per \acs{RB}\\
Channel/delay & \acs{NTN}-\acs{TDL}-A/B: 30 or 100 ns; C/D: 5 or 30 ns\\
Satellite altitude & 600 km\\
\acs{UE} elevation/speed & 10--89$^\circ$; 0--120 km/h\\
Pre-compensation error & Uniform on $[-300,300]$ Hz\\
Oscillator offset & Uniform on $[-100,100]$ Hz\\
\acs{SNR}/samples & 0:2:20 \acs{dB}; 11,264 per condition\\
Receiver correction & UL-\acs{NTN}-1: enabled; UL-\acs{NTN}-2: disabled\\
\bottomrule
\end{tabular}
\end{table}

\subsection{Baseline Models and Configurations}\label{sec:experimental_models}
The benchmark includes \acs{HELENA}, \acs{HELENA}-\ac{MHSA}, \ac{SRCNN}, ChannelNet, \ac{EDSR}, \ac{AttRNet}, \ac{ProEsNet}, \ac{LSiDNN}-48, \ac{CEViT}, \ac{MDELAN}-\acs{SISO}, and \ac{CHEA}. \acs{HELENA} follows~\cite{HELENA2025}; the other estimators follow their respective published architectures~\cite{soltani2019deep,maruyama2021,Gao2025,Zhang2024,Sharma2024,CEViT2023,MDELAN2026,CHEA2026}, adapted to the common uplink grid and pilot pattern.

ChannelNet and \acs{EDSR} retain their original depths but reduce the feature width from 64 to 32 filters to lower inference latency with limited impact on accuracy. \acs{MDELAN}-\acs{SISO} uses the tuned configuration from~\cite{HELENANTN2027}. Each model is trained once per receiver condition.

Classical references are \ac{LS} with linear interpolation, practical \ac{DM-RS} estimation, and global and profile-aware \ac{LMMSE}~\cite{HELENANTN2027}. \ac{LS} and \ac{LMMSE} use pilot-location \ac{LS} estimates; the practical estimator uses the noisy grid and known reference symbols. \ac{LMMSE} uses statistics fitted on the training and validation partitions and nominal \ac{SNR}, while profile-aware selection additionally uses the true multipath profile. Deployment timing covers learned models.

\subsection{Architectural Variants of HELENA and ANNA}
We test two \acs{HELENA} input representations at different attention capacities. All \acs{HELENA} controls and \acs{ANNA} ablations use pilot \ac{LS} observations without auxiliary features. \emph{Sparse-grid \acs{HELENA} (S)} retains the original full-grid convolutional feature extractor and output-level \ac{LS} residual connection. \emph{Packed-pilot \acs{HELENA} (P)} replaces this feature extractor with \acs{ANNA}'s width-one, 32-channel convolution and \ac{ReLU}, followed by a linear projection to 64 features per token. The attention, normalization, squeeze-and-excitation, and reconstruction stages remain unchanged. The sparse \ac{LS} grid is reconstructed from packed pilots for output residual addition.

Both representations use $n_h=4$ attention heads. The per-head key dimension $d_k$ is reduced from 64 to 32, 16, or 8. Configurations are named S-H$n_h$K$d_k$ and P-H$n_h$K$d_k$, with original \acs{HELENA} serving as the S-H4K64 reference.

For \acs{ANNA}, we vary convolution width $w$, channel count $C$, and aggregation depth $L$ from the reference configuration $(C,L,w)=(32,2,5)$. W1 sets $w=1$ to process each \ac{RB} independently. W3 and W9 change the kernel width to three and nine, reducing or extending the receptive field from nine to five or seventeen \acp{RB}, respectively. C16 halves the channel count to reduce feature capacity without changing the receptive field. L1 uses one aggregation layer instead of two, reducing the receptive field to five \acp{RB}. Comparing L1 with W3 therefore tests different aggregation depths at the same receptive-field extent. All variants retain convolution--\ac{ReLU} aggregation without residual connections.


\section{Performance Evaluation}\label{sec:results}
\begin{table*}[!t]
\centering\small
\caption{Accuracy, complexity, and embedded inference cost.}
\label{tab:ul_combined}\label{tab:ul_accuracy}\label{tab:ul_deployment}
\setlength{\tabcolsep}{1.5pt}
\begin{tabular*}{\textwidth}{@{\extracolsep{\fill}}lllrrrrrrrr@{}}
\toprule
& & & \multicolumn{2}{c}{Complexity} & \multicolumn{2}{c}{Mean \acs{NMSE} (\acs{dB})} & \multicolumn{4}{c}{UL-\acs{NTN}-1 deployment only}\\
\cmidrule(lr){4-5}\cmidrule(lr){6-7}\cmidrule(lr){8-11}
& Estimator & Input type & \shortstack{Params\\($10^3$)} & \shortstack{FLOPs\\($10^6$)} & UL-\acs{NTN}-1 & UL-\acs{NTN}-2 & \shortstack{Mean\\(ms)} & \shortstack{\acs{P99}\\(ms)} & \shortstack{$>0{,}5$ ms\\(\%)} & \shortstack{Energy\\(mJ/sample)}\\
\midrule
\multirow{6}{*}{\rotatebox{90}{ANNA}} & Reference & Packed LS & 21,936 & 2,215 & -19,06 & -20,17 & 0,33 & 0,37 & 0,048 & 1,81\\
 & W1 & Packed LS & 13,744 & 1,380 & -14,28 & -14,62 & 0,25 & 0,28 & 0,003 & 1,40\\
 & W3 & Packed LS & 17,840 & 1,798 & -18,34 & -19,29 & 0,31 & 0,34 & 0,015 & 1,70\\
 & \cellcolor{blue!8}W9 & \cellcolor{blue!8}Packed LS & \cellcolor{blue!8}30,128 & \cellcolor{blue!8}3,051 & \cellcolor{blue!8}-19,45 & \cellcolor{blue!8}-20,77 & \cellcolor{blue!8}0,33 & \cellcolor{blue!8}0,35 & \cellcolor{blue!8}0,006 & \cellcolor{blue!8}1,79\\
 & \cellcolor{blue!8}C16 & \cellcolor{blue!8}Packed LS & \cellcolor{blue!8}8,576 & \cellcolor{blue!8}0,855 & \cellcolor{blue!8}-18,75 & \cellcolor{blue!8}-19,72 & \cellcolor{blue!8}0,30 & \cellcolor{blue!8}0,32 & \cellcolor{blue!8}0,011 & \cellcolor{blue!8}1,64\\
 & \cellcolor{blue!8}L1 & \cellcolor{blue!8}Packed LS & \cellcolor{blue!8}16,784 & \cellcolor{blue!8}1,692 & \cellcolor{blue!8}-18,40 & \cellcolor{blue!8}-19,34 & \cellcolor{blue!8}0,25 & \cellcolor{blue!8}\textbf{0,27} & \cellcolor{blue!8}0,006 & \cellcolor{blue!8}\textbf{1,38}\\
\midrule
\multirow{9}{*}{\rotatebox{90}{HELENA}} & Reference & Sparse LS & 116,290 & 83,931 & -18,91 & -20,12 & 0,72 & 1,00 & 100 & 3,39\\
 & \acs{MHSA} & Sparse LS & 114,162 & 83,921 & -18,36 & -19,05 & 0,70 & 0,96 & 100 & 3,22\\
 & \cellcolor{blue!8}S-H4K32 & \cellcolor{blue!8}Sparse LS & \cellcolor{blue!8}83,138 & \cellcolor{blue!8}80,569 & \cellcolor{blue!8}-19,56 & \cellcolor{blue!8}-20,68 & \cellcolor{blue!8}0,68 & \cellcolor{blue!8}0,95 & \cellcolor{blue!8}100 & \cellcolor{blue!8}3,15\\
 & S-H4K16 & Sparse LS & 66,562 & 78,888 & -19,10 & -20,47 & 0,67 & 0,97 & 100 & 3,13\\
 & S-H4K8 & Sparse LS & 58,274 & 78,048 & -18,94 & -20,32 & 0,73 & 0,99 & 100 & 3,32\\
 & P-H4K64 & Packed LS & 93,120 & 9,327 & -16,69 & -16,98 & 0,46 & 0,50 & 1,172 & 2,49\\
 & P-H4K32 & Packed LS & 59,968 & 5,965 & -16,62 & -16,95 & 0,45 & 0,49 & 0,531 & 2,41\\
 & P-H4K16 & Packed LS & 43,392 & 4,284 & -16,59 & -17,05 & 0,45 & 0,49 & 0,478 & 2,42\\
 & P-H4K8 & Packed LS & 35,104 & 3,443 & -16,66 & -16,85 & 0,48 & 0,52 & 4,548 & 2,59\\
\midrule
\multirow{9}{*}{\rotatebox{90}{DL-based}} & \acs{SRCNN} & Interp. LS & 8,129 & 276,935 & -15,58 & -15,85 & 0,78 & 1,05 & 100 & 5,53\\
 & ChannelNet-F32 & Interp. LS & 177,506 & 6012,186 & -18,08 & -18,86 & 2,41 & 2,48 & 100 & 24,07\\
 & \acs{EDSR}-B8/F32 & Half-grid LS & 194,817 & 1675,918 & -18,52 & -19,45 & 1,57 & 1,64 & 100 & 11,90\\
 & \cellcolor{blue!8}\acs{AttRNet} & \cellcolor{blue!8}Pilot LS & \cellcolor{blue!8}103,186 & \cellcolor{blue!8}344,793 & \cellcolor{blue!8}\textbf{-20,19} & \cellcolor{blue!8}\textbf{-21,51} & \cellcolor{blue!8}1,34 & \cellcolor{blue!8}1,51 & \cellcolor{blue!8}100 & \cellcolor{blue!8}9,30\\
 & \acs{ProEsNet} & Pilot LS & 4,012 & 10,084 & -18,22 & -18,99 & 1,75 & 1,83 & 100 & 10,63\\
 & \acs{LSiDNN}-48 & Pilot LS & 878,880 & 1,741 & -16,90 & -17,31 & 0,26 & 0,29 & 0,012 & 1,48\\
 & \acs{CEViT} & Interp. LS + SI & 204,416 & 129,930 & -19,03 & -20,08 & 0,80 & 0,96 & 100 & 3,49\\
 & \acs{MDELAN}-\acs{SISO} & PA/DA-LS + pos. & 1,396 & 17,645 & -14,32 & -14,54 & 0,94 & 1,28 & 100 & 6,24\\
 & \acs{CHEA} & Pilot-symbol LS & 36,290 & 12,883 & -18,98 & -20,30 & 1,39 & 1,62 & 100 & 8,19\\
\midrule
\multirow{4}{*}{\rotatebox{90}{Classical}} & \acs{LS}+\acs{LI} & (see Sec.~\ref{sec:experimental_models}) & --- & --- & -4,07 & -4,11 & --- & --- & --- & ---\\
 & Practical \acs{DM-RS} & (see Sec.~\ref{sec:experimental_models}) & --- & --- & -14,59 & -15,24 & --- & --- & --- & ---\\
 & G-\acs{LMMSE} & (see Sec.~\ref{sec:experimental_models}) & --- & --- & -19,34 & -20,81 & --- & --- & --- & ---\\
 & PA-\acs{LMMSE}$^*$ & (see Sec.~\ref{sec:experimental_models}) & --- & --- & -20,61 & -22,71 & --- & --- & --- & ---\\
\bottomrule\end{tabular*}
\par\smallskip\begin{minipage}{0.98\textwidth}\scriptsize
Shading marks the UL-\acs{NTN}-1 accuracy--\acs{P99} Pareto front. Bold denotes the lowest error, \acs{P99}, or energy among learned estimators.
Packed LS groups pilots by \acsp{RB}. Sparse LS is zero-filled. Half-grid LS is $306\times7$. SI denotes nominal \acs{SNR}, mobility Doppler, and delay spread. PA/DA-LS denotes pilot-/data-aided estimates, and pos. denotes frequency/time coordinates. G/PA denotes global/profile-aware \acs{LMMSE}. $^*$Uses the true channel profile.
\end{minipage}
\end{table*}

\subsection{Evaluation Setup}
We evaluate estimation accuracy under both receiver conditions using 1,716 held-out samples per condition. Learned models are deployed with TensorRT~10.3 in \ac{FP16} on a 10~W Jetson Orin NX with batch size one. Latency and energy measurements use UL-\acs{NTN}-1 checkpoints. Following the receiver-processing allocation in~\cite{HELENA2025,Damjancevic2021}, we adopt a 0.5~ms \ac{P99} estimator budget, corresponding to one slot at 30~kHz \ac{SCS}.

For each model, 100 repeated benchmark runs use 500 warm-up and 5,000 timed calls, with balanced randomized model ordering and five-second idle guards. Host-to-host latency includes memory transfers, inference, synchronization, and output copying from prepared inputs. Percentiles are computed from the pooled 500,000 calls, and \ac{P99} is also assessed per run. Energy is measured separately over ten sustained-workload runs, each lasting at least ten seconds and 5,000 calls after warm-up. Energy per sample is total module-input energy divided by processed samples. Mean \ac{NMSE} equally averages per-\ac{SNR} values in linear scale before conversion to \acs{dB}. The models and datasets will be released as open source\footnote{\url{https://github.com/miguelhdo/anna-helena-ntn}}.

\subsection{Estimation Accuracy and Architectural Choices}
Table~\ref{tab:ul_combined} reports accuracy and complexity for the 24 learned configurations and four classical estimators. Across both receiver conditions, deployed \ac{FP16} mean \ac{NMSE} differs from the \ac{FP32} reference by at most 0.0066~\acs{dB}. Reference \acs{ANNA} achieves $-19.058$~\acs{dB} in UL-\acs{NTN}-1 and $-20.171$~\acs{dB} in UL-\acs{NTN}-2, compared with $-18.907$ and $-20.118$~\acs{dB} for \acs{HELENA}. It therefore retains comparable accuracy with 81.1\% fewer parameters and 37.88$\times$ fewer profiled \ac{FLOPs}. Paired values below refer to UL-\acs{NTN}-1 and UL-\acs{NTN}-2, respectively.

The \acs{HELENA} variations reveal different accuracy costs. Reducing attention capacity to S-H4K32 improves mean \ac{NMSE}, while further reductions increase error. Replacing full-grid feature extraction with packed-pilot projection has a larger impact, increasing error by 2.22--2.94~\acs{dB} in UL-\acs{NTN}-1 and 3.14--3.74~\acs{dB} in UL-\acs{NTN}-2 at matched attention settings.

The \acs{ANNA} ablations show the benefit of local frequency context. W1 processes each \ac{RB} independently and reaches only $-14.277$/$-14.619$~\acs{dB}. W3 improves accuracy, while W9 reaches $-19.453$/$-20.769$~\acs{dB} with a receptive field of 17~\acp{RB}. The wider kernels provide both greater context and more parameters. C16 and L1 trade accuracy for reduced feature capacity or depth. Among the other learned estimators, \acs{AttRNet} achieves the lowest mean \ac{NMSE}, improving on \acs{ANNA}-W9 by 0.741/0.740~\acs{dB}, while \acs{CEViT} and \acs{CHEA} remain close to reference \acs{ANNA}. Global \ac{LMMSE} also provides competitive accuracy, lying between reference \acs{ANNA} and W9 in UL-\acs{NTN}-1 and within 0.039~\acs{dB} of W9 in UL-\acs{NTN}-2. Profile-aware \ac{LMMSE} achieves lower error using the true multipath profile and nominal \ac{SNR}.

\begin{figure}[!t]
    \centering
    \includegraphics[width=\columnwidth]{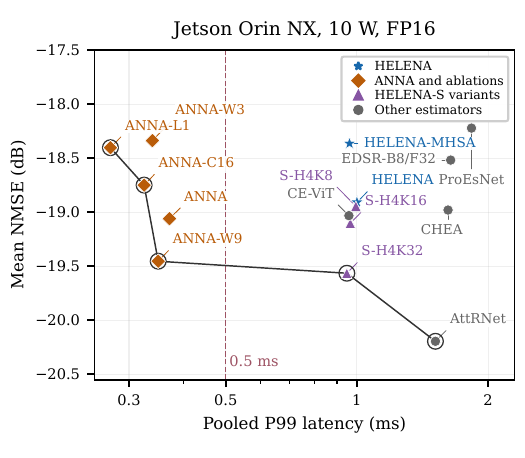}
    \caption{UL-\acs{NTN}-1 accuracy--latency trade-off (\acs{NMSE} below $-17.5$~\acs{dB}, \acs{P99} below 2.2~ms). Circles mark Pareto front; dashed line marks the 0.5~ms budget.}
    \label{fig:ul_pareto}
\end{figure}

\subsection{Deployment Performance}
Reference \acs{ANNA} achieves 0.372~ms pooled \ac{P99}, meeting the budget and providing a 2.69$\times$ speed-up over \acs{HELENA}'s 1.0004~ms. Attention-reduced \acs{HELENA} remains above budget at 0.949--0.995~ms. Packed-pilot variants reach 0.491--0.520~ms: P-H4K32 and P-H4K16 meet the pooled \ac{P99} target, but exceed it in 10 and eight of the 100 runs, respectively, and incur the accuracy loss observed above. Neither simplification route achieves \acs{ANNA}'s accuracy--latency trade-off.

W9 improves accuracy while reaching 0.350~ms \ac{P99}. L1 prioritizes latency, reaching 0.272~ms and a 3.68$\times$ speed-up at a 0.506~\acs{dB} accuracy penalty relative to \acs{HELENA}. C16 provides an intermediate trade-off at 0.325~ms and $-18.748$~\acs{dB}. Among the other estimators, \acs{LSiDNN}-48 reaches 0.287~ms but has higher error than L1, whereas \acs{AttRNet}'s accuracy advantage requires 1.515~ms \ac{P99}.

Figure~\ref{fig:ul_pareto} summarizes the accuracy--\ac{P99} trade-off. L1, C16, W9, S-H4K32, and \acs{AttRNet} form the Pareto front, with the \acs{ANNA} variants below the adopted budget. S-H4K32 and \acs{AttRNet} extend the front toward higher accuracy at greater latency. All six \acs{ANNA} configurations meet the 0.5~ms \ac{P99} target in every benchmark run. Reference \acs{ANNA} and W9 achieve pooled 99.9th-percentile latencies of 0.407 and 0.371~ms, respectively. Individual calls exceed 0.5~ms in 0.048\% of reference \acs{ANNA} measurements and 0.006\% for W9.

Reference \acs{ANNA} and W9 consume 1.814 and 1.791~mJ per sample at the module input, respectively, 46.4\% and 47.1\% less than \acs{HELENA}'s 3.387~mJ. All \acs{ANNA} variants consume less energy than the other learned baselines except \acs{LSiDNN}-48 at 1.481~mJ. L1 achieves the lowest consumption overall at 1.378~mJ per sample.

These results show that reducing input size or attention capacity alone does not achieve the desired accuracy--latency trade-off. The packed-pilot \acs{HELENA} variants lose accuracy, while attention reduction leaves inference above the adopted budget. \acs{ANNA} combines compact pilot processing with local frequency interaction, retaining comparable accuracy at lower latency and energy. Within this design, W9 favors accuracy, while L1 offers lower inference cost, allowing the configuration to reflect receiver requirements.

\section{Conclusion and Future Work}\label{sec:conclusion}
We proposed \ac{ANNA} for \ac{NTN} uplink \ac{CE} on regenerative payloads. It combines \ac{RB}-aligned pilot projection, convolutional neighborhood aggregation, and shared linear reconstruction, deferring dense-grid expansion until the output. It achieves comparable mean \ac{NMSE} to \acs{HELENA} in both receiver conditions with 81.1\% fewer parameters. In \ac{FP16} deployment on a 10~W Orin NX, \acs{ANNA} achieves a 2.69$\times$ speed-up and consumes 46.4\% less total module energy per sample than \acs{HELENA}. Its pooled \ac{P99} latency is 0.372~ms, and all 100 benchmark runs meet the adopted 0.5~ms \ac{P99} target. W9 and L1 offer further accuracy--latency trade-offs, showing how compact pilot processing and neighborhood aggregation support efficient embedded \ac{CE}.

Future work will evaluate link-level \ac{BER}, \ac{BLER}, and throughput, and extend the study to \ac{MIMO}, different \ac{RB} allocations, and pilot configurations. We will also investigate \ac{INT8} quantization and platform-specific execution optimizations.

\bibliographystyle{IEEEtran}
\bibliography{main}
\end{document}